\documentclass[aps,prb,twocolumn,showpacs,tightenlines]{revtex4}
\usepackage{graphics}
\usepackage{epsfig}
\begin{document}
\title{\Large  \bf{Hidden quantum pump effects in quantum coherent rings} }
\author{M. Moskalets$^{1}$ and M. B\"uttiker$^2$}
\affiliation{
         $^1$Department of Metal and Semiconductor Physics,\\
        National Technical University "Kharkov Polytechnical Institute",
        Kharkov, Ukraine\\
        $^2$D\'epartement de Physique Th\'eorique, Universit\'e de Gen\`eve,
        CH-1211 Gen\`eve 4, Switzerland\\
}

\date\today

\begin{abstract}
Time periodic perturbations of an electron system on a ring are 
examined. For small frequencies periodic small amplitude 
perturbations give rise to side band currents which 
in leading order are inversely proportional to the frequency. 
These side band currents compensate 
the current of the central band such that to leading 
order no net pumped current is generated.  In the non-adiabatic limit, 
larger pump frequencies can lead to resonant excitations: 
as a consequence a net pumped current 
arises. We illustrate our results for a one channel ring with a 
quantum dot whose barriers are modulated parametrically. 
\end{abstract}


   \maketitle 

\small

\section{Introduction}
\label{i}
\indent

Recently Switkes et al. \cite{SMCG99} demonstrated that 
a phase-coherent mesoscopic sample subjected to 
a cyclic two parameter perturbation can produce a directed current.
Of interest is a quantum pump effect which arises solely due to
quantum-mechanical interference and dynamical breaking of time-reversal invariance.
This pump effect can be elegantly expressed with the help of scattering
matrices \cite{Brouwer98,VAA01} in close analogy to ac-transport
in mesoscopic structures \cite{BTP94,MPPB02}.
Research in this field is currently very active. 
We refer the reader only to a few recent related works, 
on charge quantization \cite{quant}, 
the role of dephasing \cite{MB01,CB02}, heat generation 
by pumps \cite{AEGS00,MB02,WW02} 
and noise \cite{AEGS00,MB02,PVB02} 
and the transition from adiabatic to non-adiabatic 
transport \cite{MBstrong02,trans}. 
Additional related problems addressed are for example, 
adiabatic pumping in hybrid super-conducting normal structures \cite{super}, 
Cooper pair pumps \cite{hekking},  
spin-pumping \cite{spin},  
the magnetic field symmetry \cite{symm}. 
For an extensive list of references to earlier work we refer the reader
to Refs. \onlinecite{MPPB02} and \onlinecite{MB02}.
Since a pump current results even in the limit of a slow
variation of the pump parameters and in the limit of small 
amplitude variation of these parameters, the system under consideration 
is still close to its equilibrium state. Parametric pumping 
provides therefore an approach to examine near-equilibrium  properties 
of the system which can not be obtained by conductance measurements. 

\begin{figure}
  \vspace{3mm}
  \centerline{
   \epsfxsize8cm
   \epsffile{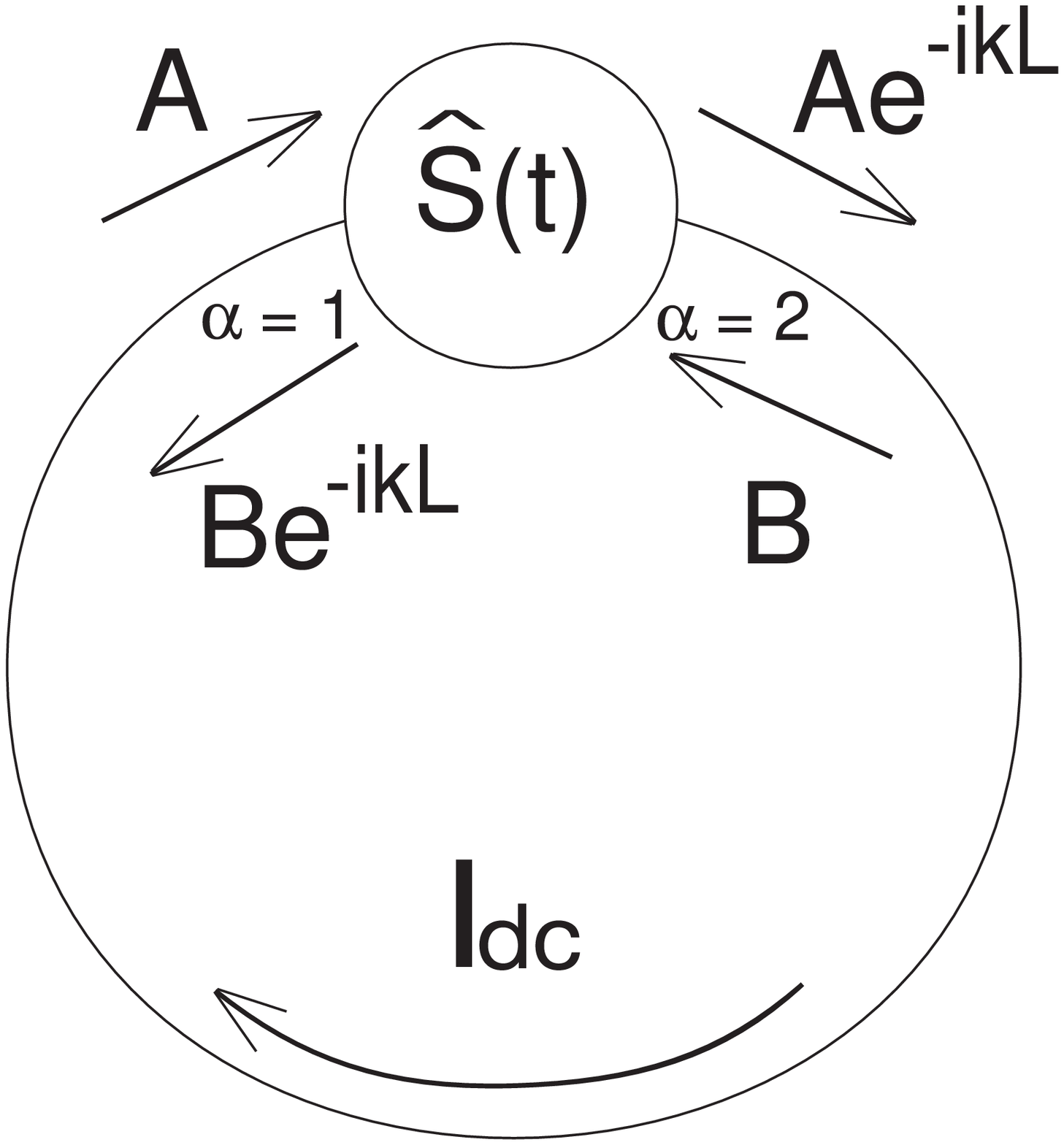}
             }
  \vspace{3mm}
  \nopagebreak
  \caption{
A quantum dot with scattering matrix $\hat S$ is embedded in a 
one-dimensional ring. 
The Greek letter $\alpha$ numbers the scattering channels.
$\Psi(x)= A e^{ik(x-L)} + Be^{-ikx}$ is an electron wave function.
$x=0$ and $x=L$ correspond to the right and left sides 
of a dot, respectively.
If the scattering matrix depends on time $\hat S = \hat S(t)$ then
a directed circulating current $I_{dc}$ can arise. 
}
\label{fig1}
\end{figure}

Of most interest have been open systems, like the one investigated in   
Ref.\onlinecite{SMCG99}, for which the particle spectrum is continuous. 
As a consequence even at small frequencies $\omega \to 0$ 
pumped electron current is in a strictly quantum-mechanical sense {\it non-adiabatic}. 
This is because an oscillating scatterer induces transitions 
between electron states separated by one or several modulation quanta
$\hbar\omega$ and in a system with continuos spectrum this implies 
that the system is driven out of equilibrium at arbitrarily small 
frequencies. 

In this work we consider an electron system on a ring.
At sufficiently low temperatures, when the phase coherence length 
is much larger than the diameter of the ring, such a system exhibits 
a spectrum which is essentially discrete. It is well known that in such 
a system purely static breaking of 
the left-right (L-R) symmetry by a magnetic field (by an 
Aharonov-Bohm flux) can generate at {\it equilibrium} 
directed currents. These are the well known "persistent currents" \cite{Kulik70,BIL83} 
which were observed experimentally \cite{LDDB90,CWBKGK91,MCB93}.
Here we are interested in the generation of directed currents 
in such a ring due to pumping in the absence of a symmetry breaking 
magnetic field.  
Instead of the magnetic field the left-right symmetry is 
dynamically broken by the oscillating scatterer.
It is the purpose of this work to investigate in more detail the physical 
processes underlying quantum pumping of electrons along a ring.
We use the Floquet function representation for an electron wave function 
in the periodically driven systems.
We show that if the potentials oscillate out of phase 
(the time reversal symmetry is {\it dynamically} broken)
then each component of the Floquet state carries a current.
This is the reason why a net circulating current (a pumped current)
arises even if the potentials oscillate with a small amplitude. 
In the present paper we concentrate on this small amplitude limit.

Note that employing the Floquet approach allows us
to consider the pump effect in closed systems 
and in open systems \cite{MBstrong02} on the same footing. 
The scattering approach has been useful in 
the discussion of parametric pumping in open systems 
and we demonstrate here its applicability to pumping in closed 
systems. Other approaches, more closely related to a linear response 
approach are also possible \cite{Cohen02,ARZ}. 
The Floquet approach used here, allows us to show that a pumped current exists 
even in the absence of near degeneracies, at pumping amplitudes which are so small 
that they do not lead to level crossings.

For small frequencies and small amplitudes, we find that 
a parametric oscillation of the scatterer generates a Floquet 
state with a current both in the main branch of the wave function 
as well as in its side bands. Interestingly, these currents 
are in leading order {\it inversely} proportional to frequency. 
Moreover, to leading order they compensate 
one another. Thus at small frequencies and small amplitudes pumping 
generates currents at different energies within a Floquet state 
without (to leading order) generating a net total pumping current. 
We term this phenomena 
a "hidden pump effect". We present an analytical (exact) discussion 
of this effect. We also support the analytical discussion with a numerical 
calculation of a specific model and present results for the non-adiabatic 
case.

\section{A general Floquet scattering approach}
\label{gfa}
\indent

We consider a one-channel ring of length $L$ with embedded scatterer
(a quantum dot) of a small size $w\ll L$ as shown in  Fig.\ref{fig1}. 
We suppose that the scattering matrix of the quantum dot 
oscillates in time with frequency $\omega$. 
Then according to the Floquet theorem we can write 
the single-electron wave functions $\Psi(x,t)$ as follows

\begin{equation}
\label{Eq1}
\Psi_E(x,t) = e^{-iEt/\hbar}\sum^\infty_{n=-\infty}\psi_n(x)
e^{-in\omega t}.
\end{equation}

\noindent
Here $E$ is the Floquet energy.
Each Floquet state  can be occupied by only one electron
(because of the Pauli principle)
and thus the wave function $\Psi_{E}$ must be normalized

\begin{equation}
\label{Eq2}
\begin{array}{l}
 \frac{1}{\cal{T}}\int\limits_0^{\cal{T}} dt
\int\limits_0^{L_r} dx |\Psi_E|^2
\equiv \sum\limits_n \int\limits_0^{L_r} dx |\psi_n|^2
= 1.
\end{array}
\end{equation}

\noindent 
Here ${\cal{T}} = 2\pi/\omega$.
For the ring problem under consideration
we choose functions $\psi_n(x)$ in the following form

\begin{equation}
\label{Eq3}
\psi_n(x) = A_n e^{ik_n (x-L)} + B_n e^{-ik_n x}.
\end{equation}

\noindent
Here $k_n = \sqrt{2m_eE_n/\hbar^2}$ 
with $Re[k_n]\geq  0$ and $Im[k_n]\geq 0$. Furthermore, 
$E_n = E + n\hbar\omega$ is the side band energy. 

The coefficients $A_n$ and $B_n$ with different index $n$ 
are coupled through the cyclic boundary conditions 
at the oscillating scatterer. 
We express them in terms of
the Floquet scattering matrix $\hat S_F$ 
relating the incoming waves $A_m, B_m$ 
to outgoing ones $A_ne^{-ik_nL}, B_ne^{-ik_nL}$ (see Fig.\ref{fig1}). 
The matrix element $S_{F,\alpha\beta}(E_n,E)$ defines 
the quantum mechanical amplitude ${\cal A}_{\alpha\beta}(E_n,E)$
for the particle coming from the channel $\beta$ with energy $E$ to be
scattered into the channel $\alpha$ after the emission ($n<0$) or
the absorption ($n>0$) of $n$ energy quanta $\hbar\omega$:

\begin{equation}
\label{Eq4}
{\cal A}_{\alpha\beta}(E_n,E) = \sqrt{\frac{k}{k_n}} 
S_{F,\alpha\beta}(E_n,E).
\end{equation}

\noindent
Numbering the scattering channels as it is shown in Fig.\ref{fig1}
we find that the boundary conditions imply:

\begin{equation}
\label{Eq5}
\begin{array}{c}
A_ne^{-ik_nL} = \sum\limits_{m=-\infty}^{\infty} 
\sqrt{\frac{k_{m}}{k_n}}  \\
\times\left[ A_{m} S_{F,21}(E_n,E_{m}) 
+ B_{m} S_{F,22}(E_n,E_{m}) \right], \\
\ \\
B_ne^{-ik_nL} = \sum\limits_{m=-\infty}^{\infty}  
\sqrt{\frac{k_{m}}{k_n}}  \\
\times\left[ A_{m} S_{F,11}(E_n,E_{m}) 
+ B_{m} S_{F,12}(E_n,E_{m}) \right]. 
\end{array}
\end{equation}

Thus we obtain an infinite system of uniform linear equation 
for the coefficients $A_n$ and $B_n$. To have a nontrivial solution
the corresponding (infinite range) determinant must be equal to zero.
This (dispersion) equation determines the allowed values 
of the Floquet energy $E^{(l)}$ (where $l$ is an integer) and
the corresponding set of coefficients 
$A_n^{(l)}$ and $B_n^{(l)}$ of the Floquet wave
function Eqs.(\ref{Eq1}) and (\ref{Eq3}).

In practice only a finite number of sidebands have to 
be taken into account. For instance, in the case of weak pumping
(when the corresponding potentials oscillate with small amplitudes) 
only the first side bands are essential. In this case
we can put $n=0,\pm 1$ in
Eq.(\ref{Eq1}) and Eq.(\ref{Eq5})
reduces to a system of only six equations.

Next we consider the current carried by 
the particular Floquet state $\Psi_{E^{(l)}}$. 
We will concentrate on
the time averaged (dc) current $I_{dc}$. 
To this end we integrate the quantum mechanical current
(in what follows the star denotes complex conjugation)

\begin{equation}
\label{Eq6}
I[\Psi] = - \frac{e\hbar}{m_e} {\rm Im}\left[
\Psi\frac{\partial \Psi^*}{\partial x}
\right],
\end{equation}

\noindent
over the time period ${\cal{T}} = 2\pi/\omega$

\begin{equation}
\label{Eq7}
I_{dc}^{(l)} = 
\frac{1}{\cal{T}}\int\limits_0^{\cal{T}} dt I[\Psi_{E^{(l)}}],
\end{equation}

\noindent
and obtain

\begin{equation}
\label{Eq8}
\begin{array}{l}
 I_{dc}^{(l)} = \sum\limits_{E_n^{(l)} > 0} I^{(l)}_n, \\
\ \\
I^{(l)}_n = \frac{e\hbar}{m_e}
k_n^{(l)}\left(|A_n^{(l)}|^2 - |B_n^{(l)}|^2
 \right).
\end{array}
\end{equation}

\noindent
Here we have restricted the summation over the propagating modes 
($E_{n}^{(l)} > 0$) only, since the bounded states ($E_n^{(l)}<0$) 
do not contribute to the current.

To solve Eq.(\ref{Eq5}) and calculate the current Eq.(\ref{Eq8})
one needs to know the Floquet scattering matrix $\hat S_F$. 
Therefore in what follows we consider some particular cases.
Furthermore, we concentrate on the current carried by 
a single electron state. 
To find the full circulating current we have to sum Eq.(\ref{Eq8})
over all the occupied levels in the ring.

\section{Currents generated by an oscillating energy independent scatterer}
\label{eis}
\indent

The Floquet scattering matrix depends in general on 
both the energy of the incident carriers and the energy of the 
exiting carriers [see Eq. (\ref{Eq4})]. 
However, for frequencies which are small compared 
to the typical energy over which the scattering matrix 
varies significantly, the 
Floquet scattering 
matrix can be expressed in terms of one energy argument only, 
i.e. in terms of the on-shell scattering matrix of the 
stationary problem  \cite{MBstrong02}.
To be definite we assume that the stationary scattering matrix depends 
on two parameters $\hat S$ = $\hat S(p_1, p_2)$
which oscillate with frequency $\omega$
and with phase lag $\varphi$:

\begin{equation}
\label{Eq9}
\begin{array}{l}
p_1 = p_{01} + 2p_{11}\cos(\omega t + \varphi/2); \\
\ \\
p_2 = p_{02} + 2p_{12}\cos(\omega t - \varphi/2).
\end{array}
\end{equation}

\noindent
Then the corresponding Floquet scattering matrix $\hat S_F$
can be expressed in terms of the Fourier coefficients of $\hat S$ 
as follows \cite{MBstrong02} 

\begin{equation}
\label{Eq10}
\begin{array}{c}
\hat S_F(E_n,E) = \hat S_{n\omega}, \\
\ \\
\hat S_{n\omega} = \frac{\omega}{2\pi} 
\int\limits_{0}^{2\pi/\omega} dt e^{in\omega t} 
\hat S(p_1(t),p_2(t)).
\end{array}
\end{equation}
Note that we restrict our consideration to scatterers
which, in the absence of the time-dependent perturbations,  
are time reversal invariant: $S_{12} = S_{21}$. In particular, 
this implies that there is no stationary circulating current.

In what follows we concentrate on the small amplitude

\begin{equation}
\label{Eq11}
    p_{1i} \ll p_{0i}, ~~~i = 1,2.
\end{equation}

\noindent
and the small frequency case:

\begin{equation}
\label{Eq12}
\hbar\omega\ll \Delta,
\end{equation}

\noindent
where $\Delta$ is the level spacing of the ring
(near the level $E^{(l)}$ under consideration). 
In the small amplitude limit Eq.(\ref{Eq11})
only the side bands with $n=\pm 1$ are excited. 
Their intensities are proportional to the following Fourier coefficients

\begin{equation}
\label{Eq13}
\hat S_{\pm\omega} = 
p_{11}e^{\mp i\frac{\varphi}{2}}\frac{\partial\hat S}{\partial p_{01}} 
+ p_{12}e^{\pm i\frac{\varphi}{2}}\frac{\partial\hat S}{\partial p_{02}}, 
\end{equation}

\noindent
and, hence, they are small.
Therefore to find the Floquet eigen energies $E^{(l)}$ 
in the lowest order in the oscillating amplitudes $p_{1i}$
it is sufficient to consider only the two equations of Eq.(\ref{Eq5})
which correspond to the coefficients $A_0$ and $B_0$. The other 
amplitudes can be taken to be zero. 
The resulting dispersion equation 
coincides with that of the time-independent problem,
i.e., the dispersion equation 
for a ring with a static quantum dot having the scattering matrix
$\hat S$ = $\hat S(p_{01},p_{02})$. 
In terms of the components of the scattering matrix 
the dispersion equation reads

\begin{equation}
\label{Eq14}
D[k]\equiv
\left(
e^{-ikL} - S_{12}\right)^2 - S_{11}S_{22} = 0.
\end{equation}

\noindent
This equation defines the allowed set of eigenvectors
$k^{(l)}$ and the corresponding Floquet eigen energies
$E^{(l)} = (\hbar k^{(l)})^2/(2m_e)$.

Apparently only the main component ($n=0$) of the Floquet state
$\Psi_E$ Eq.(\ref{Eq1}) is subject to constructive interference 
on the ring. In contrast, 
the side bands with energies 
$E_{\pm 1}^{(l)} = E^{(l)} \pm \hbar\omega$
are subject to destructive interference.
That is another reason why their intensities are small.
Note that this is correct if the frequency $\omega$ is small enough
Eq.(\ref{Eq12}) and the side bands are not close to other eigen energies
of the ring.

We choose some particular energy level $E^{(l)}$ and consider the currents
$I^{(l)}_n$ carried by the different components ($n = -1, 0, +1$) 
of the Floquet state. 
In what follows we drop the upper index $^{(l)}$.
To calculate these currents, induced by the oscillating scatterer, 
we have to take into account the processes of absorption and emission
of energy quanta $\hbar\omega$. 
In the weak amplitude limit these processes being unimportant
for the eigen energy problem nevertheless lead to a noticeable change 
in the wave function: the resulting 
different components of the Floquet state give rise 
to circulating currents. 

First we calculate the current $I_{+1}$ [see Eq.(\ref{Eq8})]
associated with the upper side band of the Floquet state.  
To find the current in the lowest order in oscillating amplitudes 
$p_{1i}$ we use Eq.(\ref{Eq5}) 
and express the coefficients $A_{+1}$ and $B_{+1}$ 
in terms of $A_0$ and $B_0$:

\begin{equation}
\label{Eq15}
\begin{array}{r}
A_{+1}e^{ik_{+1}L}\left(e^{-ik_{+1}L} 
- S_{12}\right) - B_{+1}S_{22} \\
= A_{0}e^{ikL}S_{12,+\omega} + B_{0}S_{22,+\omega}; \\
\ \\
-A_{+1}e^{ik_{+1}L}S_{11} 
+ B_{+1}\left(e^{-ik_{+1}L} - S_{12}\right) \\
= A_{0}e^{ikL}S_{11,+\omega} + B_{0}S_{12,+\omega}.
\end{array}
\end{equation}

\noindent
We neglect other side bands and use the following equation 
(which holds in the zero order in $p_{1i}$)

\begin{equation}
\label{Eq16}
\begin{array}{l}
B_0 \approx A_0e^{ikL}\frac{e^{-ikL} - S_{12}}{S_{22}} \\
\ \\
= A_0e^{ikL}\frac{S_{11}}{e^{-ikL} - S_{12}}. \\
\end{array}
\end{equation}

\noindent
In addition to determine the coefficient $A_0$ we use
the normalization condition

\begin{equation}
\label{Eq17}
|A_0|^2 + |B_0|^2 \approx \frac{1}{L}.
\end{equation}

\noindent
Note that for any solution of Eq.(\ref{Eq14})
($k = k^{(l)}$) 
the equation (\ref{Eq16}) gives $|A_0|^2$ = $|B_0|^2$.

Substituting Eq.(\ref{Eq16}) into Eq.(\ref{Eq15}) we get

\begin{equation}
\label{Eq18}
\begin{array}{r}
A_{+1}e^{ik_{+1}L} = A_{0}e^{ikL}
\frac{e^{-ikL} - S_{12}}{D[k_{+1}]} \\
\ \\
\times 
\left\{ \Pi[S_{22}]\left(e^{-ik_{+1}L} - S_{12}\right) \right.\\
\ \\
\left.
+\Pi[S_{11}]\left(e^{-ikL} - S_{12}\right)
\right\}, 
\ \\
\ \\
B_{+1} = B_{0}
\frac{e^{-ikL} - S_{12}}{D[k_{+1}]}~~~~~~~~~~~~~~~~ \\
\ \\
\times 
\left\{ \Pi[S_{11}]\left(e^{-ik_{+1}L} - S_{12}\right) \right.\\
\ \\
\left.
+\Pi[S_{22}]\left(e^{-ikL} - S_{12}\right)
\right\},
\end{array}
\end{equation}

\noindent
where

\begin{equation}
\label{Eq19}
\Pi(S_{ii}) = \frac{S_{ii,+\omega}}{S_{ii}}
+\frac{S_{12,+\omega}}{e^{-ikL} - S_{12}}.
\end{equation}

The function $D[k_{+1}]$ entering the denominators in Eq.(\ref{Eq18})
is defined in Eq.(\ref{Eq14}). 
At small frequency $\omega$ [see Eq.(\ref{Eq12})]
we can expand $k_{+1}\approx k + \omega/v$ 
(here $v = \hbar k/m_e$ is an electron velocity)
and obtain (since $D[k]=0$)

\begin{equation}
\label{Eq20}
D[k_{+1}] \approx -2i\frac{\omega}{\omega_0}e^{-ikL}
\left(e^{-ikL} -S_{12}\right).
\end{equation}

\noindent
Here $\omega_0 = v/L$. This equation is of the lowest order in
the ratio $\omega/\omega_0$. 

The function $D[k_{+1}]$ describes the effect of a destructive 
interference on the side band ($n = +1$).
On the other hand the main component ($n = 0$) is subject to 
constructive interference. 
Hence the smaller the frequency $\omega$ 
(and, thus, the closer the energy $E_{+1} = E + \hbar\omega$ 
of the side band to the eigen energy $E$)
the weaker is the destructive interference.
As a result the amplitude of a wave function 
$\psi_{+1}\sim\frac{1}{D[k_{+1}]}$
and, correspondingly, the current $I_{+1}$ {\it increases} 
with decreasing $\omega$. 

Such a dependence $A_{+1}, B_{+1}, I_{+1} \sim \omega^{-1}$
holds while the wave function $\psi_{+1}$ is still small.
At extremely small $\omega$
the effect of the oscillating parameters 
has to be taken into account exactly (not perturbatively).
Numerical calculations (see \cite{MBstrong02}) show 
that in this case higher side bands are excited.
The limiting case of such small frequencies will be considered elsewhere.

Using Eq.(\ref{Eq18}) we calculate the current 
$I_{+1}$ $\sim$ $Re[(A_{+1}$ + $B_{+1})(A_{+1}$ - $B_{+1})^{*}]$.
After simple manipulations we can express the current in terms of
quantities $\Pi_{(\pm)} = $ $\Pi[S_{22}] \pm$ $\Pi[S_{11}]$.
To find $\Pi_{(+)}$ we use the dispersion equation (\ref{Eq14}) 
and take into account that the eigenvector $k$ depends on 
the parameters: $k = k(p_{01},p_{02})$.
Differentiating Eq.(\ref{Eq14}) with respect to either $p_{01}$ or $p_{02}$
we obtain 
\begin{equation}
\label{Eq21}
\Pi_{(+)} = \frac{-2iLk_{+\omega}e^{-ikL}}{e^{-ikL}-S_{12}}.
\end{equation}
\noindent
Here $k_{+\omega}$ is defined in the same fashion as 
$\hat S_{+\omega}$ Eq.(\ref{Eq13}).
To find $\Pi_{(-)}$ we use the identity $|S_{11}|^2$ = $|S_{22}|^2$
and get
\begin{equation}
\label{Eq22}
\Pi_{(-)} = 2i\theta_{+\omega}
\end{equation} 
\noindent
where $\theta_{+\omega}$ is a Fourier coefficient of the phase  
\begin{equation}
\label{Eq23}
 \theta = \frac{i}{2} ln\left(\frac{S_{11}}{S_{22}}\right).
\end{equation}
\noindent 
Note that the (real) phase $\theta$ characterizes the asymmetry
in the reflection of particles incident from the left ($S_{11}$) and
from the right ($S_{22}$). This asymmetry is due to the spatial asymmetry
of a quantum dot.
Finally the current $I_{+1}$ reads as follows

\begin{equation}
\label{Eq24}
\begin{array}{l}
I_{+1} = I_{\omega}\sin{\varphi}, \\
\ \\
I_{\omega} = \frac{ev}{L} 
\frac{p_{11}p_{12}}{\hbar\omega}
\left(
  \frac{\partial\theta}{\partial p_{01}}\frac{\partial E}{\partial p_{02}} 
- \frac{\partial\theta}{\partial p_{02}}\frac{\partial E}{\partial p_{01}} 
\right).
\end{array}
\end{equation}
\noindent 
Here 
$E = E(p_{01},p_{02})$
is an eigen energy for the static problem. 

From Eq.(\ref{Eq24}) it follows that 
if two parameters oscillate out of phase $\varphi\neq 0,\pi$
then a dc current $I_{+1}$ arises. This is similar to the 
case of an open quantum cavity 
where an oscillating scatterer pumps a dc current between
external reservoirs \cite{Brouwer98}.
However the dependence on the frequency $\omega$ is 
strikingly different. This indicates that this is a new phenomena 
which is specific for closed systems. 

Like the pumped current in open systems \cite{MBstrong02} 
the current $I_{+1}$ under considerations is due to dynamical breaking 
of the time reversal symmetry ($\varphi\neq 0,\pi$) 
by the oscillating scatterer.
Note that there is another necessary condition for the existence of 
dynamically generated dc currents
which is general for open and closed systems: The varying parameters 
must affect the spatial asymmetry of the scatterer, i.e.,
$\partial\theta/\partial p_{i}\neq 0$.

In addition the general conditions mentioned above,
there exists a particular condition which is
specific for the ring problem under consideration.
The current $I_{\omega}$ depends on the sensitivity 
of eigen energies to the varied parameters. 
If we have (accidentally) $\partial E/\partial p_i = 0$,  
then the current is zero.
In particular, in the limit of an extremely small scatterer ($w\to 0$)
one can classify the eigenstates $\psi_{E^{(l)}}(x)$ in a ring 
according to their spatial symmetry. 
If the scatterer is at $x = 0$ then for the (anti-) symmetric states 
we have $\psi(x) = (-)\psi(-x)$. 
The antisymmetric states are insensitive to the presence of 
a small quantum scatterer (because $\psi_{(anti)}(x=0) \approx 0$)
and their energies do not depend on $p_{i}$. 
Thus in such a case 
the antisymmetric states do not exhibit the pump effect discussed
here. 

Now we calculate the currents carried by the other components of
the Floquet state.
To obtain the current $I_{-\omega}$ we need to replace
$\omega\to -\omega$ and $\varphi\to -\varphi$.
From Eq.(\ref{Eq24}) it follows that $I_{-\omega} = I_{+\omega}$.
A similar calculation shows that $I_{0} = -(I_{+\omega}+I_{-\omega})$.
Thus within the approximation used the full circulating current
\begin{equation}
\label{Eq25}
I_{dc} = I_{0} + I_{-1} + I_{+1}
\end{equation}
is zero.
\noindent 
Because it is impossible to measure the current carried by only 
one side band (only a full current is a measurable quantity) 
we call the effect under consideration a {\it "hidden pump effect"}.

The disappearance of the full circulating current is a consequence of the symmetry
between the side bands corresponding to absorption $\psi_{+1}$ and
to emission $\psi_{-1}$ of modulation quanta $\hbar \omega$. 
This symmetry can be broken if the energy of one of the side bands 
(either $E_{+1}$ or $E_{-1}$) lies close to another eigen energy in the ring. 
In this case a net circulating current can arise: $I_{dc}\neq 0$. 
Thus if the adiabaticity condition Eq.(\ref{Eq12})
is violated then the effect under consideration can be measured.
In the next section we present the results of a numerical calculation
confirming such a conclusion. 

It is important to note that the pumped currents discussed here 
are inversely proportional to the frequency, 
$I\sim \omega^{-1}$. Therefore even in the non-adiabatic regime 
the pumping effect discussed here 
differs strongly from the pump effect in open systems 
where $I\sim \omega$ \cite{SMCG99,Brouwer98}.
Nevertheless in addition to the {\it "hidden pump effect"}
there exists a true pump effect \cite{MBaqpc02}
(for which the circulating current is proportional to $\omega$) 
and which is fully analogous to the pump effect in open systems.
To obtain this current one needs to carry out the expansions 
in powers of the frequency to a higher order then is done here. 
We will present the results of such a higher order expansion 
in a separate work \cite{MBaqpc02}.

\section{Currents generated by an oscillating double barrier}
\label{db}
\indent

In this section we use a simple model for the quantum dot
to calculate numerically the currents generated in the ring. 
We consider both the adiabatic case Eq.(\ref{Eq12}) and   
the nonadiabatic case 
to confirm the conclusions of the previous section.

We model a quantum dot by
two delta-function potentials separated by the distance $w$ 
and choose the strength of these potentials $V_{1}$ and $V_{2}$
as varying parameters.

Appendix A gives an exact solution of the model under consideration. 
In the numerical calculations 
we use the units $2m_e = \hbar = e = 1$ and put 
$\varphi_1 = -\varphi_2 = \varphi/2$, where $\varphi$ is the phase lag.
Note that at $\varphi = 0$ two potentials oscillate in phase.
We concentrate on the case of opaque 
($V_{1(2)} \gg k\hbar^2/m_e$) closely placed ($w\ll L_r$) 
barriers which correspond to a quantum dot only weakly coupled to a ring. 
\begin{figure}
  \vspace{3mm}
  \centerline{
   \epsfxsize9cm
   \epsffile{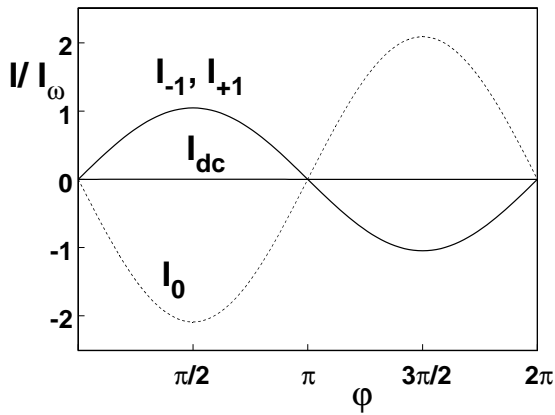}
             }
  \vspace{3mm}
  \nopagebreak
  \caption{
Adiabatic case.
The off resonance currents carried by the main component $I_0$ 
and by the first two side bands $I_{+1} = I_{-1}$ 
of the Floquet state 
are given as a function of the phase lag 
$\varphi = \varphi_1 - \varphi_2$.
The full current $I_{dc} = \sum_n I_n$  
and currents carried by the higher side bands are vanishingly small.
The currents are given in units of $I_{\omega}$ Eq.(\ref{Eq27}).
The parameters of the oscillating double barrier are: 
$V_{01} = V_{02} = 400$; 
$V_{11} = V_{12} = 0.04$.
$k = 9.546$;
$\omega = 0.1$;
$L_r = 10\pi$; 
$w = \pi/20$.
We use the units: $2m_e = \hbar = e = 1$. 
}
\label{fig2}
\end{figure}
\begin{figure}
  \vspace{3mm}
  \centerline{
   \epsfxsize9cm
   \epsffile{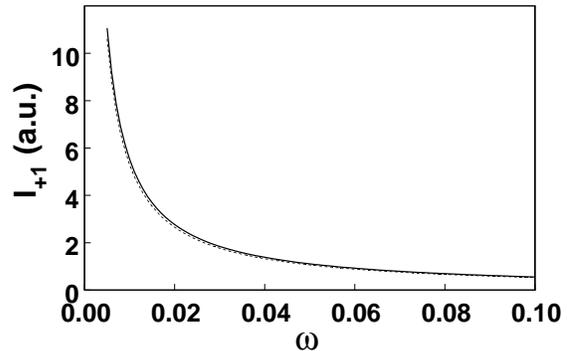}
             }
  \vspace{3mm}
  \nopagebreak
  \caption{
Adiabatic case.
The dependence of the current $I_{+1}$ at $\varphi = \pi/2$
on the frequency $\omega$: (i) numerical calculations (solid line);
(ii) estimation according to Eq.(\ref{Eq27}) (dashed line).
The parameters are the same as in Fig.\ref{fig2}.
}
\label{fig3}
\end{figure}

\subsection{Adiabatic case}
\label{ac}
\indent

We consider small frequencies Eq.(\ref{Eq12}) 
and calculate the current carried by some energy level $E$ in the ring
which is far from any energy level in the dot (off resonance case).
In this case the oscillating potentials have only
a weak effect on the wave function amplitudes which are mainly
determined by the interference due to the ring geometry \cite{MBring02}.

In Fig.\ref{fig2} we depict the dependence of the currents 
carried by the main component $I_0$ 
and by the first two side bands $I_{+1}$ and $I_{-1}$ 
of the Floquet state on the phase lag
$\varphi$ between the oscillating potentials $V_{11}$ and $V_{12}$
(see Appendix A).
The full current $I_{dc}$ (which is vanishingly small) is shown as well.

At small amplitudes of the oscillating potentials 
only the lowest side bands ($n = \pm 1$) are excited.
In the case under consideration the symmetry 
between the absorption and the emission is preserved. 
As a result both side bands $E_{+1}$ and $E_{-1}$ 
have the same amplitudes and carry the same currents:$I_{+1} = I_{-1}$.
The current $I_0$ is twice as large as $I_{\pm 1}$ 
and opposite to $I_{\pm 1}$.
This is in agreement with the previous section.

To estimate the magnitude $I_{\omega}$ of a current 
we note that far from the resonance and at small frequencies
the double barrier scattering matrix 
can be considered as energy independent \cite{MBstrong02}.
Thus we can use Eq.(\ref{Eq24}) with the scattering matrix
of a double barrier potential:
\begin{equation}
\label{Eq26}
 \hat S(p_{01},p_{02}) = \frac{e^{ikw}}{\Delta}
\left( 
\begin{array}{cc}
\xi + 2\frac{p_{02}}{k}\sin(kw) & 1 \\
\ \\
1 & \xi + 2\frac{p_{01}}{k}\sin(kw) 
\end{array}
\right).
\end{equation}
\noindent
Here 
$p_{0i} = V_{0i}m_e/\hbar^2$ (i = 1,2);
$\xi$ = $(1-\Delta)e^{-ikw}$;
$\Delta$ = $1 + \frac{p_{01}p_{02}}{k^2}(e^{2ikw} -1)$ 
+ $i\frac{p_{01}+p_{02}}{k}$.

Far from the resonance (and for $k\ll p_{0}$) we get: 
$\partial k/\partial p_{i} \sim k/(2p_{0}^2L)$;
$\partial\theta/\partial p_{01} = -
\partial\theta/\partial p_{02}\sim k/(2p_{0}^2)$.
Substituting these estimates into Eq.(\ref{Eq24}) we obtain
\begin{equation}
\label{Eq27}
I_{\omega} \approx 
2eT\frac{\omega_0^2}{\omega}\left(\frac{p_1}{k}\right)^2.
\end{equation}
\noindent 
Here $T=k^4/(4p_0^4)\ll 1$ 
is an off resonance probability for tunneling 
through the double barrier potential. 
We put $p_{01} = p_{02} = p_{0}$ and $p_{11} = p_{12} = p_{1}$.

The above equation 
[see also Eq.(\ref{Eq24})]
predicts that 
the amplitude of a current carried by the individual side band 
scales as $\omega^{-1}$. This is illustrated in Fig.\ref{fig3}.
We can see that the analytical results Eq.(\ref{Eq24}) and Eq.(\ref{Eq27})
are in a good agreement with the results of numerical calculations.
\begin{figure}[t]
  \vspace{3mm}
  \centerline{
   \epsfxsize9cm
   \epsffile{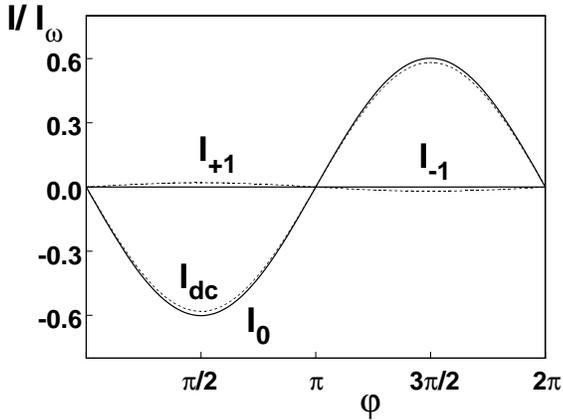}
             }
  \vspace{3mm}
  \nopagebreak
  \caption{
Nonadiabatic case.
The currents carried by the individual side bands 
$I_{-1}$, $I_{0}$, $I_{+1}$
and the full circulating current $I_{dc} = \sum_n I_n$ 
are given as a function of the phase lag 
$\varphi = \varphi_1 - \varphi_2$ close to the resonance.
The currents are given in units of $I_{\omega}$ Eq.(26)
with $T$ being an actual transmission probability
at energy $E = 89.84$ 
and $\omega$ replaced by $\omega_r - \omega$, where
$\omega_r = (E_r - E)/\hbar = 2.34$ is a resonance frequency.
The transmission resonance through the quantum dot occurs at
$E_r = 92.18$. 
The width of the resonance is $\Gamma = 0.13 \ll \omega$.  
The side band $E_{+1} \equiv E + \hbar\omega = 92.15$
is close to the transmission resonance 
whereas $E$ and $E_{-1}$ are far from the resonance.
The parameters of the oscillating double barrier are: 
$V_{01} = V_{02} = 400$; 
$V_{11} = V_{12} = 0.04$.
$k = 9.4785$;
$\omega = 2.31$;
$L_r = 2\pi$; 
$w = \pi/9.75$.
We use the units: $2m_e = \hbar = e = 1$. 
}
\label{fig4}
\end{figure}

\begin{figure}[t]
  \vspace{3mm}
  \centerline{
   \epsfxsize9cm
   \epsffile{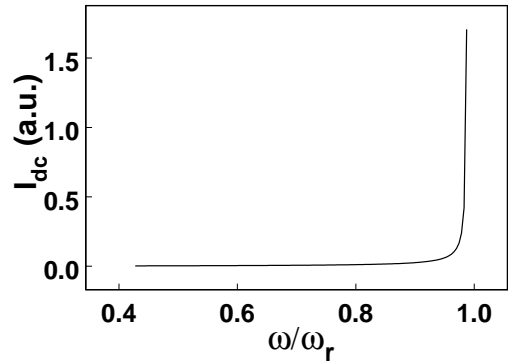}
             }
  \vspace{3mm}
  \nopagebreak
  \caption{
Nonadiabatic case.
The dependence of the circulating current $I_{dc}$ 
on the frequency $\omega$ at $\varphi = -\pi/2$.
The parameters are the same as in Fig.\ref{fig4}.
}
\label{fig5}
\end{figure}

\subsection{Nonadiabatic case}
\label{nac}
\indent

In this subsection 
we consider the conditions 
under which a net circulating current 
arises in a ring with oscillating double barrier potential: $I_{dc}\neq 0$.
As we pointed out already if the symmetry between the emission 
and the absorption holds the full current
carried by the Floquet state is zero.
However if this symmetry is destroyed then 
a net circulating current can arise.
In particular the symmetry between the absorption and the emission 
is destroyed if one of the side bands (say $E_{+1}$) 
is close to the transmission resonance through the quantum dot. 
In this case tunneling after absorbing of a modulation quantum $\hbar\omega$
dominates over tunneling after emitting $\hbar\omega$.

In Fig.\ref{fig4} we depict the dependence of the net pumped current 
$I_{dc}$ and currents carried by the individual side bands 
$I_{0}$, $I_{+1}$, and $I_{-1}$
on the phase difference $\varphi = \varphi_1 - \varphi_2$.
At given parameters the side band $E_{+1} = E + \hbar\omega$ 
is close to the transmission resonance through the quantum dot
whereas the basic energy level $E$ and the side band 
$E_{-1} = E - \hbar\omega$ are out of resonance:
$E - E_r\gg \Gamma$, where $E_r$ is the resonance energy and
$\Gamma\approx \frac{\hbar^2}{2m_e}\frac{2k^3}{wp_0^2}$ 
is the width of the transmission resonance.
When $\omega$ approaches the resonance frequency 
$\hbar\omega_r = E_r - E$ then the circulating
current increases. This is illustrated in  Fig.\ref{fig5}.

From Fig.\ref{fig4} we see that 
in the nonadiabatic case under consideration 
the circulating current $I_{dc}$
is carried, in fact, by the main component $\psi_{0}$ of the Floquet state.
The mechanism which generates this current is as follows.
An electron (mainly staying at the level with energy $E$) 
can absorb an energy $\hbar\omega$
and tunnel resonantly ($E_{+1}\approx E_{r}$) through 
the quantum dot. 
After tunneling an electron emits an energy $\hbar\omega$ and 
returns to its original energy level. 
Because of the phase lag $\varphi$ between the oscillating potentials
$V_{1}(t)$ and $V_2(t)$ the amplitudes for tunneling to the left
and to the right are different. Thus an electron 
tunnels more frequently to one side and a net circulating current arises.
Note that both the emission and the absorption affect the current.
However if $E < E_r$ then 
only the processes where the emission follows the absorption 
(or vice versa if $E > E_r$)
contribute to the current $I_{dc}$.

\section{Conclusion}
\label{c}
\indent

We have considered a quantum pump  
in a mesoscopic ring with embedded quantum dot.
If two (or more) parameters affecting the scattering properties of 
a quantum dot change periodically but out of phase then 
a circulating dc current can be generated.
We have emphasized the small frequency and small 
amplitude case when the oscillations do not affect the spectrum
(the positions of energy levels are unchanged)
but rather break dynamically the time reversal symmetry  of the system.
We have examined the features of parametric pumping which 
are specific for 
closed doubly connected systems. 
The resulting pumped current is due to a 
competition between exciting an electron system by an oscillating scatterer 
and interference due to the ring geometry.

The effect of an oscillating scatterer
on an electron wave function is twofold.
On the one hand, because of the oscillations
the system is nonstationary and the electron is in a Floquet state.
This state is characterized by the set of substates (side bands) 
with energies 
$E_{n} = E + n\hbar\omega$, $n = 0, \pm1, \pm2, \dots$.
On the other hand, if the time reversal symmetry is broken then
each substate carries a current. 
Interference results in an 
unusual $\omega^{-1}$ dependence of the current amplitude
carried by the individual side band.

A main feature of the pump effect under consideration 
is that in the adiabatic case the currents carried by the different
sub states of a given Floquet state compensate each other.
Therefore we term this effect "hidden".
Nevertheless in the nonadiabatic case the current carried by the main
component of the Floquet state dominates 
and a net circulating current $I_{dc}$ appears.

We again emphasize that, in addition to the effect considered here,  
there exists a usual adiabatic pump effect with 
a circulating current proportional to $\omega$.
This effect, 
analogous to the pump effect in open systems, 
will be considered elsewhere \cite{MBaqpc02}.

\begin{acknowledgments}
This work is supported by the Swiss National Science Foundation.
\end{acknowledgments}

\appendix

\section{A ring with two oscillating barriers. }
\indent
\label{es}

In this Appendix we determine the single-particle eigen energies and
the coefficients of the corresponding Floquet wave function Eq.(\ref{Eq1})
for a ring of length $L_r$ with two oscillating 
delta function potentials separated by a distance $w$.

The electron wave functions $\Psi(x,t)$ is a solution of 
the time-dependent Schr{\"o}dinger equation

\begin{equation}
\label{Eq28}
\begin{array}{l}
i\hbar\frac{\partial\Psi(x,t)}{\partial t} = 
\hat H(x,t)\Psi(x,t), \\
\ \\
\hat H(x,t) = - \frac{\hbar^2}{2m_e}
\frac{\partial^2}{\partial x^2} + V(x,t), \\
\  \\
V(x,t) = V_{1}(t)\delta(x) 
+ V_{2}(t)\delta(x-w), \\
\ \\
V_{i}(t) = V_{0i} + 2V_{1i}\cos(\omega t + \varphi_i), i = 1,2.
\end{array}
\end{equation}

According to the Floquet theorem the wave function 
is given by Eq.(\ref{Eq1}).
In addition to Eq.(\ref{Eq3})
we define the functions $\psi_n(x)$ inside the dot as well
\begin{equation}
\label{Eq29}
\psi_n(x) = \left\{
\begin{array}{l}
a_n e^{ik_n x} + b_n e^{-ik_n x}, 0 \leq x \leq w, \\
\ \\
A_n e^{ik_n x} + B_n e^{-ik_n x}, w \leq x \leq L_r.
\end{array}
\right.
\end{equation}
\noindent

The function $\Psi_E(x,t)$ is periodic in $x$ with the period of $L_r$.
Hence  the boundary conditions at $x = 0$ and $x = w$ read as follows

\begin{equation}
\label{Eq30}
\begin{array}{c}
\Psi_{E}(L_r,t) = \Psi_{E}(0,t), \\
\ \\
\Psi_{E}(w-0,t) = \Psi_{E}(w+0,t), \\
\ \\
{\left.\frac{\partial\Psi_E(x,t)}{\partial x}\right|_{x=0}} -
{\left.\frac{\partial\Psi_E(x,t)}{\partial x}\right|_{x=L_r}} \\
\ \\
 = \frac{2m}{\hbar^2}V_{1}(t)\Psi_{E}(L_r,t),  \\
\ \\
{\left.\frac{\partial\Psi_E(x,t)}{\partial x}\right|_{x=w+0}} -
{\left.\frac{\partial\Psi_E(x,t)}{\partial x}\right|_{x=w-0}} \\
\ \\
 = \frac{2m}{\hbar^2}V_{2}(t)\Psi_{E}(w,t).  \\
\end{array}
\end{equation}

\noindent 
These boundary conditions determine the discrete set of the Floquet
eigen energies $E^{(l)}$ (where $l$ is an integer) and corresponding
Floquet eigenfunctions $\Psi_{E^{(l)}}$.

Note that because of the Floquet theorem the time-dependent 
problem with oscillating potential Eq.(\ref{Eq28}) 
is reduced to a time-independent one.
The cost we need to pay is a splitting of each energy level 
$E^{(l)}$ into a ladder $E^{(l)} + n\hbar\omega$ 
(n = 0, $\pm1$, $\pm2$,$\dots$) \cite{MBring02}.

To find the Floquet eigen energies $E^{(l)}$
we apply the method described in \cite{MBstrong02,MBring02}.
We substitute Eq.(\ref{Eq1}) and Eq.(\ref{Eq28}) into the Eq.(\ref{Eq29})
and write the result in a matrix form
\begin{equation}
\label{Eq31}
\begin{array}{l}
 \hat C_{n}(L_r,w) 
\left(
\begin{array}{l}
A_n \\
B_n
\end{array}
\right) 
- \hat C_{n}(0,w) 
\left(
\begin{array}{l}
a_n \\
b_n
\end{array}
\right) = 0, \\
\ \\
\end{array}
\end{equation}

\begin{equation}
\label{Eq32}
 \begin{array}{l}
\hat U_{0n}(L_r,w) 
\left(
\begin{array}{l}
A_n \\
B_n
\end{array}
\right) 
+  \hat U_{0n}^{*}(0,-w) 
\left(
\begin{array}{l}
a_n \\
b_n
\end{array}
\right) \\
\ \\
= -2\hat F^{(+)} \hat C_{n+1}(L_r,w) 
\left(
\begin{array}{l}
A_{n+1} \\ 
B_{n+1}
\end{array}
\right) \\
\ \\
~~~ -2\hat F^{(-)} \hat C_{n-1}(L_r,w) 
\left(
\begin{array}{l}
A_{n-1} \\
B_{n-1}
\end{array}
\right).
\end{array}
\end{equation}

\noindent 
Here we have introduced the matrices

\begin{equation}
\label{Eq33}
\hat C_n(L_r,w) = 
\left(
\begin{array}{lr}
e^{ik_nL_r} & e^{-ik_nL_r} \\
\ \\
e^{ik_nw} & e^{-ik_nw} \\
\end{array}
\right),
\end{equation}

\begin{equation}
\label{Eq34}
\hat U_{0n}(L_r,w) = 
\left(
\begin{array}{lr}
e^{ik_nL_r}(p_{01} + ik_n) & e^{-ik_nL_r}(p_{01} - ik_n) \\
\ \\
e^{ik_nw}(p_{02} - ik_n) & e^{-ik_nw}(p_{02} + ik_n) 
\end{array}
\right),
\end{equation}

\begin{equation}
\label{Eq35}
\hat F^{(\pm)} = 
\left(
\begin{array}{lr}
 p_{11}e^{\pm i\varphi_1} & 0 \\
\ \\
 0 & p_{12}e^{\pm i\varphi_2} 
\end{array}
\right),
\end{equation}

\noindent 
where the parameters are:
$p_{ji} = V_{ji}m_e/\hbar^2$, $j=0,1$, $i=1,2$.
Note that the system of equations (\ref{Eq30}) and (\ref{Eq31}) 
represents an infinite number of linear equations. 
To simplify it we introduce new matrixes $\hat X_n$  as follows

\begin{equation}
\label{Eq36}
\begin{array}{l}
 -2\hat F^{(\pm)} \hat C_{n\pm 1}(L_r.w) 
\left(
\begin{array}{l}
A_{n\pm 1} \\
B_{n\pm 1}
\end{array}
\right) 
= \hat X_{n\pm 1} 
\left(
\begin{array}{l}
A_n \\
B_n
\end{array}
\right).
\end{array}
\end{equation} 

\noindent
Here and hereafter the upper (lower) sign is for $n>0$ ($n<0$).
Substituting Eq.(\ref{Eq31}) and Eq.(\ref{Eq36}) into Eq.(\ref{Eq32})
we get the following recursive equation for $\hat X_{n}$
\begin{equation}
\label{Eq37}
 \hat X_n = 4\hat F^{(\pm)}\hat C_n(L_r,w)
\left(
\hat U_n - \hat X_{n\pm 1}
\right)^{-1}\hat F^{(\mp)}\hat C_{n\mp 1}(L_r,w),
\end{equation}
\noindent
where
$$
\hat U_n = \hat U_{0n}(L_r,w) + 
\hat U_{0n}(0,w)\hat C_n^{-1}(0,w)\hat C_n(L_r,w).
$$

The advantage of Eq.(\ref{Eq37}) is in the following.
In each particular case we need to take into account only
the limited number $|n| < n_{max}$ of side bands
and thus we can put $\hat X_{\pm(n_{max}+1)}\approx 0$.
After that we can easily calculate $\hat X_n$ at $|n| \leq n_{max}$
and can express all the coefficients
$A_{n}$, $B_{n}$  in terms of $A_0$ and $B_0$ only.
Thus we have used all the equations of the system 
Eqs.(\ref{Eq31}) and (\ref{Eq32}) except those for $n=0$.

The remaining part is a system of uniform equations for the coefficients
$a_0, b_0, A_0,$ and $B_0$. To have a nontrivial solution 
the corresponding determinant ($4\times 4$) must be equal to zero.
This condition determines the set of the allowed Floquet eigen energies
$E^{(l)} = (\hbar k^{(l)})^2/(2m_e)$ and corresponding side bands
$E^{(l)}_{n} = E^{(l)} + n\hbar\omega$. 
To calculate the coefficients entering 
the corresponding Floquet wave function 
$\Psi_{E^{(l)}}$ Eqs.(\ref{Eq1}),(\ref{Eq29}) 
we use the normalization condition Eq.(\ref{Eq2}).

\end{document}